\documentclass[manuscript]{aastex}

\shorttitle{Radiatively excited CH$^+$, CH, and DIBs}
\shortauthors{Oka et al.}

\begin{document}

\title{Anomalous Diffuse Interstellar Bands in the Spectrum of Herschel 36. II.
Analysis of Radiatively Excited CH$^+$, CH, and DIBs}

\author {Takeshi Oka\altaffilmark{1,2,3}, Daniel E. Welty\altaffilmark{1}, Sean Johnson\altaffilmark{1}, Donald G. York\altaffilmark{1,2},\\
Julie Dahlstrom\altaffilmark{4}, and L. M. Hobbs\altaffilmark{1,5}}

\affil{Department of Astronomy and Astrophysics, University of Chicago, 5640 South Ellis Avenue, Chicago, IL 60637. t-oka@uchicago.edu}

\altaffiltext{1}{Department of Astronomy and Astrophysics, University of Chicago, 5640 South Ellis Avenue, Chicago, IL 60637. t-oka@uchicago.edu}
\altaffiltext{2}{Enrico Fermi Institute, University of Chicago, Chicago, IL 60637.}
\altaffiltext{3}{Department of Chemistry, University of Chicago, 5735 South Ellis Avenue, Chicago, IL 60637.}
\altaffiltext{4}{Department of Physics and Astronomy, Carthage College 2001 Alford Park Dr., Kenosha, WI 53140}
\altaffiltext{5}{University of Chicago, Yerkes Observatory, Williams Bay, WI 53191.}

\begin{abstract}
Absorption spectra toward Herschel 36 for the \emph{\~{A}}$^1\Pi$~$\leftarrow$~\emph{\~{X}}$^1\Sigma$ transitions of CH$^+$ in the $J$~=~1 excited rotational level and the \emph{\~{A}}$^2\Delta$~$\leftarrow$~\emph{\~{X}}$^2\Pi$ transition of CH in the $J$~=~3/2 excited fine structure level have been analyzed. 
These excited levels are above their ground levels by 40.1 K and $\sim$~25.7~K and indicate high radiative temperatures of the environment, 14.6~K and 6.7~K, respectively. 
The effect of the high radiative temperature is more spectacular in some diffuse interstellar bands (DIBs) observed toward Her 36; remarkable extended tails toward red (ETR) were observed. We interpret these ETRs as due to a small decrease of rotational constants upon excitation of the excited electronic states.
Along with radiative pumping of a great many high-$J$ rotational levels, this causes the ETRs. 
In order to study this effect quantitatively, we have developed a model calculation in which the effects of collision and radiation are treated simultaneously. 
The simplest case of linear molecules is considered. 
It has been found that the ETR is reproduced if the fraction of the variation of the rotational constant, $\beta$~$\equiv~(B'-B)/B$, is sufficiently high (3~$-$~5\%) and the radiative temperature is high ($T_\mathrm{r}$~$>$~50~K).  
Although modeling for general molecules is beyond the scope of this paper, the results indicate that the prototypical DIBs $\lambda$5780.5, $\lambda$5797.1, and $\lambda$6613.6 which show the pronounced ETRs are due to polar molecules sensitive to the radiative excitation.  
The requirement of high $\beta$ favors relatively small molecules with 3-6 heavy atoms. 
DIBs $\lambda$5849.8, $\lambda$6196.0, and $\lambda$6379.3  which do not show the pronounced ETRs are likely due to non-polar molecules or large polar molecules with small $\beta$.
\end{abstract}


\section{INTRODUCTION}

Since their discovery (Heger 1922), the diffuse interstellar bands (DIBs) have been a persistent spectroscopic enigma -- representing unidentified ingredients in the physics and chemistry of the interstellar medium.  
The DIBs are ubiquitous in the ISM -- with the strongest clearly visible toward even lightly reddened stars -- though their relative strengths can vary in different sight lines (e.g., Kre{\l}owski \& Walker 1987).
Recent deep surveys have catalogued at least 540 DIBs, to equivalent width limits of several m\AA, in two sight lines with $E$($B-V$)~$\sim$~1.1 (Hobbs et al. 2008, 2009).
While the DIBs are likely molecular in origin, attempts to identify specific carrier molecules by comparing interstellar spectra with laboratory spectra of various candidates have been unsuccessful (e.g., Snow \& McCall 2006; Salama et al. 2011).
Some characteristics of the carriers may be inferred, however, from the observed absorption-line profiles.
High-resolution, high-S/N spectra have revealed substructure in the profiles of a number of the DIBs (Sarre et al. 1995; Kre{\l}owski \& Schmidt 1997; Galazutdinov et al. 2008).
That substructure was initially attributed to either rotational structure (Ehrenfreund \& Foing 1996) or isotopic substitution (Webster 1996) in large, carbon-based carrier molecules in the gas phase -- which could contain a non-negligible fraction of the available carbon.
Variations in the separations and relative strengths of the subcomponents -- presumably reflecting differences in local environmental conditions -- may be seen for a number of the DIBs (Cami et al. 2004; Galazutdinov et al. 2008).
While the variations in subcomponent separation seem to rule out isotopic substitution, modelling of the substructures as rotational contours can provide constraints on the size, shape, and rotational constants of the carrier molecules (Ehrenfreund \& Foing 1996; Schulz et al. 2000; Cami et al. 2004).
If the DIB carriers can be identified (and their behavior understood), they will provide potentially significant, widely available diagnostics of the physical conditions in the ISM.

Several potentially important clues to the behavior (and, ultimately, the identities) of the DIB carriers have recently been found in optical spectra of the heavily reddened star Herschel 36 (Her 36).
Dahlstrom et al. (2013; hereafter Paper I) discovered both absorption from excited levels of CH$^+$ and CH and very strong extended tail toward red (ETR) in the profiles of a number of the DIBs toward Her 36.
Absorption from excited interstellar CH$^+$ and CH has not been seen in any other sight line.
While the ETR can be seen for some DIBs in a few other sight lines, they are generally much weaker than those seen toward Her 36.
In this paper, we determine that the molecular excitation must be due to strong far-infrared pumping from the adjacent IR source Her 36 SE , and we propose that IR pumping is also responsible for the strong ETRs on the DIBs.
If that is the case, then some constraints may be placed on the properties of the molecular carriers of the DIBs.
In sections 2 and 3, we discuss the excitation of CH$^+$ and CH, respectively.
In section 4, we compare the observed DIB profiles with simulated profiles for linear molecules subject to both collisional and radiative excitation.
In sections 5 and 6, we discuss what can be inferred about the properties and identities of the molecules responsible for the DIBs.

\section{CH$^+$ IN THE $J$~=~1 EXCITED LEVEL AND RADIATIVE TEMPERATURE}

\subsection{Past Observations}

The spectrum of interstellar CH$^+$ was discovered by Theodore Dunham, Jr. (1937) as an unidentified absorption line at 4232.6 \AA~toward $\zeta$~Persei, $\chi^2$ Orionis, 55 Cygni, and $\chi$~Aurigae.  
Douglas \& Herzberg (1941) identified the spectrum as due to the \emph{\~{A}}$^1\Pi$~$\leftarrow$~\emph{\~{X}}$^1\Sigma$ (0,0) transition of CH$^+$ by laboratory emission spectroscopy. 
This was the first identification of a molecular ion in interstellar space.
Since then interstellar CH$^+$ has been observed through absorption at rest wavelengths (in air) 4232.5~\AA~(0, 0), 3957.7~\AA~(1, 0), and 3745.3~\AA~(2, 0) toward a great many stars, demonstrating the ubiquity of CH$^+$ in the diffuse interstellar medium (e.g. Gredel et al. 1993; Crawford 1995; Gredel 1997).  
Nearly all of the more than 250 detections of CH$^+$ absorption lines so far have been of the $R$(0) transitions of CH$^+$ in the ground $J$~=~0 rotational level.  
The sole exception is the circumstellar absorption toward HD 213985 reported by Bakker et al. (1997).  
Because of the very fast $J$~=~1~$\rightarrow$~0 spontaneous emission (See Section 2.3.), the excited $J$~=~1 level is not sufficiently populated for detection in ordinary diffuse clouds.  
Highly excited CH$^+$ lines have been observed in visible emission from the Red Rectangle (Balm \& Jura 1992; Hall et al. 1992; Bakker et al. 1997; Hobbs et al. 2004) and in far-infrared emission in the planetary nebula NGC 7027 (Cernicharo et al. 1997), the Orion Bar photon-dominated region (Habart et al.2010; Naylor et al. 2010) and the disc around Herbig Be star HD 100546 (Thi et al. 2011). 
Model calculations for the emissions observed in NGC 7027 and the Orion Bar have recently been published (Godard \& Cernicharo 2013; Nagy et al. 2013).

\subsection{CH$^+$ Spectrum toward Herschel 36}

The \emph{\~{A}}$^1\Pi$~$\leftarrow$~\emph{\~{X}}$^1\Sigma$ (0,0) and (1,0) bands of CH$^+$ from the archive of the Fiberfed Extended Range Optical Spectrograph (FEROS) with a resolution of 48,000 (Kaufer et al. 1999) toward Herschel 36 and 9 Sgr are compared in Fig.~1.  
Both stars are members of the young cluster NGC 6530.  
The 3 arcmin separation between the two sight lines corresponds to a linear separation of about 1.3~pc at the roughly 1.5~kpc distance to the cluster. 
The $R$(1) and $Q$(1) absorptions of CH$^+$ in the $J$~=~1 excited rotational level, 40.1~K above the ground $J$~=~0 level, are clearly seen for both vibronic bands only toward Her 36.
The $R$(2) and $Q$(2) lines for the $J$~=~2 level, 120.3~K above the ground level, are not detectable.  
The non-detection of the $J$~=~1 lines toward 9 Sgr suggests that the excited CH$^+$ is found in a small region near Herschel 36 (Paper I).

Observed transitions, with oscillator strength $f$, observed (heliocentric) wavelength in air $\lambda_\mathrm{air}$, measured equivalent width of the absorption $W_\lambda$, H\"{o}nl-London factor $S$, derived CH$^+$ column density in the $J$ level $N$($J$), and excitation temperature $T_\mathrm{ex}$ are listed in Table 1. 
For weak lines, the column densities could be derived from $W_\lambda$ by using $N$~=~($mc^2$/$\pi$$e^2$)($W_{\lambda}$/$f\lambda^2$), where $f$~=~(8$\pi^2m$/3$he^2$)$\nu$$\mu_t$$^2$ is the oscillator strength and $\mu_t$ is the transition dipole moment (e.g., Spitzer 1978).
The FITS6P profile-fitting program (Welty et al. 2003) was used to fit stronger lines (accounting for any saturation in the absorption) and to fit multiple lines from a given species.  
A $b$-value of 1.2~-~1.6~km~s$^{-1}$ yielded consistent $N$(0) for the (0,0), (1,0), and (2,0) bands of CH$^+$; we adopt $N$(0)~=~9.4~$\times$~10$^{13}$~cm$^{-2}$ and $N$(1)~=~1.8~$\times$~10$^{13}$~cm$^{-2}$.  
The excitation temperature $T_\mathrm{ex}$ was derived from $N$($J$)/$N$(0)~=~(2$J$~+~1)exp(-$E_J$/$kT_\mathrm{ex}$).
Uncertainties are 1$\sigma$; upper limits for undetected lines correspond to 3$\sigma$.

The values of $W_\lambda$ should be identical for $R$(1) and $Q$(1) (which have the same H\"{o}nl-London factors), and the values of $N$($J$) and $T_\mathrm{ex}$ should be identical for the (0,0) and (1,0) bands.  
The slight differences in those measured and derived values reflect our measurement uncertainties.  
We proceed in the following using the excitation temperature of 14.6~K determined from the three CH$^+$ bands.

\subsection{High Radiative Temperature of 14.6~K}

The observed high excitation temperature of $T_\mathrm{ex}$~=~14.6~K is exceptional, in view of the very short life time of CH$^+$ in the $J$~=~1 level due to the fast spontaneous emission $J$~=~1~$\rightarrow$~0.  
The energy difference between the $J$~=~1 and 0 levels of CH$^+$ has been known since the work by Douglas \& Herzberg (1941) but was recently measured directly by Amano (2010) to be 835,137.504 MHz, which is equivalent to 27.86~cm$^{-1}$ and 40.08~K.  
The permanent dipole moment of CH$^+$ has been calculated by \emph{ab initio} theory to be $\mu$~=~1.656~Debye (Ornellas \& Machado 1986), 1.679 Debye (Follmeg et al. 1987), and 1.804 Debye (Sun \& Freed 1988).  
If we use $\mu$~=~1.7~Debye, we have the Einstein coefficient $A_\mathrm{10}$~=~(64$\pi^4\nu^3$/3$hc^3$)$\mu^2$/3~=~0.0065~s$^{-1}$, corresponding to the lifetime of the $J$~=~1 level of 150~s. 
The critical density for this decay is estimated to be on the order of $n$(H)~$\sim$~$A_{10}$/$k_\mathrm{L}$~$\sim$~7~$\times$~$10^6$~cm$^{-3}$, where $k_\mathrm{L}$~$\sim$~1~$\times$~$10^{-9}$~cm$^3$~s$^{-1}$ is the Langevin rate constant (Anicich \& Huntress 1986).  
At such a high density, carbon will be primarily in the form of CO and any CH$^+$ would be rapidly destroyed by collisions with molecular hydrogen.  
The number density in the diffuse interstellar medium where CH$^+$ abounds is typically on the order of 10$^2$~cm$^{-3}$ or less, which is insufficient to maintain CH$^+$ in the $J$~=~1 level by collisions.  
This is the reason why CH$^+$ in the $J$~=~1 level has not previously been observed in ordinary sight lines.

The high excitation temperature of 14.6~K must be due to radiative pumping in the environment, most likely by dust emission from the nearby extended infrared source Her~36~SE (Goto et al. 2006).  
The intensity of radiation must depend on the radial position of the sight line and therefore what determines the rotational distribution of CH$^+$ must be a complicated weighted average of the radiative temperature $<T_\mathrm{r}>$ along the CH$^+$ column.   
If the whole column of CH$^+$ is irradiated by the dust emission, the average radiative temperature $<T_\mathrm{r}>$  due to dust emission would be equal to the excitation temperature, $<T_\mathrm{ex}>$~=~14.6~K.
$<T_\mathrm{r}>$ needs to be higher if only part of the CH$^+$ column is irradiated, but we believe this is not the case since CH$^+$ in the $J$~=~2 level is not detected.  
$<T_\mathrm{r}>$ cannot be much higher than the upper limit on the excitation temperature for $J$~=~2, $<T_\mathrm{ex}>$~$<$~22.9~K.

The logic used in this subsection to determine the radiative temperature $<T_\mathrm{r}>$ is basically the same as that missed in McKellar's near discovery of the 2.73 K primordial blackbody radiation (McKellar 1941, 1949) but later used by Thaddeus and Clauser (1966) and Meyer and Jura (1984) to determine the temperature of the blackbody radiation from the $R$(0) and $R$(1) lines of CN.  
The cases of CH$^+$ and CH are cleaner than that of CN because their $J$~=~1~$\rightarrow$~0 and $J$ = 3/2 $\rightarrow$ 1/2 spontaneous emissions (respectively) are orders of magnitude faster.

\section{CH SPECTRUM TOWARD HERSCHEL 36}

The effect of the high radiative temperature is also observed for the \emph{\~{A}}$^2\Delta$~$\leftarrow$~\emph{\~{X}}$^2\Pi$ (0,0) transition of CH toward Her 36.  
The absorption spectrum has previously been observed only for CH in the blended $\lambda$ doublet in the ground $J$~=~1/2~F$_2$ fine structure level through the $R_2$(1/2) transition.  
Toward the special sightline of Her 36, however, we see a CH line from the excited fine structure level $J$~=~3/2~F$_1$, 25.6~K above the ground level, through the $R_1$(3/2) transition as shown in Fig.~2.  
To the best of our knowledge this is the first time that this line has been observed in absorption (Paper I). 
[The line has been observed in emission from the Red Rectangle (Hobbs et al. 2004).]  
The observed CH equivalent widths and derived column densities are listed in Table 1.  
Data for the weaker \emph{\~{B}}$^2\Sigma^-$~$\leftarrow$~\emph{\~{X}}$^2\Pi$ (0,0) transitions, which confirm the $J$~=~1/2 column density derived from \emph{\~{A}}$^2\Delta$~$\leftarrow$~\emph{\~{X}}$^2\Pi$ (0,0), are also listed.

With the large permanent dipole moment of $\mu$~=~1.46~Debye (Phelps \& Dalby 1966; see also Kalemos et al. 1999), and large energy separation of 536,795.678 - 532,721.333~MHz (Amano 2000), the spontaneous emission $J$~=~3/2~F$_1$~$\rightarrow$~$J$~=~1/2~F$_2$ is orders of magnitude faster than collisional pumping in a diffuse cloud.  
We thus conclude, as for the case of CH$^+$, that the excitation temperature is equal to the effective radiative temperature $<T_\mathrm{r}>$ of the environment of the observed CH column. 
The small column density ratio $N$(3/2)/$N$(1/2)~=~0.043, determined by using the H\"{o}nl-London factors of Luque \& Crosley (1996), gives $<T_\mathrm{r}>$~=~6.7(1)~K.  
This is considerably lower than 14.6~K for CH$^+$.  
Perhaps this is not surprising since it is well known that CH and CH$^+$ can exist in different parts of a cloud.  
Although the velocity of CH determined from the unresolved $\lambda$ doublet lines is in agreement with that of CH$^+$, suggesting that CH and CH$^+$ are in the same cloud, the lower temperature for CH suggests that only a part ($\sim$~35\%) of the CH column is contained in the high $<T_\mathrm{r}>$ region.
Hereafter we simply write $T_\mathrm{r}$ for $<T_\mathrm{r}>$ for brevity.

\section{DIFFUSE INTERSTELLAR BANDS TOWARD HERSCHEL 36}

\subsection{The Remarkable Extended Tail toward Red (ETR) Observed in Some DIBs}

The dust emission which populates the $J$~=~1 excited rotational level of CH$^+$ and the $J$~=~3/2 excited fine structure level of CH affects the velocity profiles of some DIBs in a more spectacular manner.  
Six of the DIBs observed toward Her 36 are compared with those in the ordinary, adjacent sight line toward 9~Sgr in Fig.~3.  
For all six DIBs, the velocity profiles toward 9~Sgr (in red) are more or less symmetric with respect to the peak position.
The three DIBs $\lambda$5780.5, $\lambda$5797.1, and $\lambda$6613.6 toward Herschel 36 (in black) shown on the left of Fig.~3, however, are very asymmetric, with a huge tail extending to the red.  
We will call this Extended Tail toward Red as ETR.  
There are some indications of ETR for the three DIBs on the right of Fig.~3 also, but we have not been able to explain them quantitatively by the mechanism we consider in this paper.  
They may be due to vibrational hot bands, as discussed in Section 5.3.  
In any case, the qualitative difference between the variations of the three DIBs on the left and those on the right of Fig.~3 is obvious.  
The broad deep absorption observed for 9 Sgr near $\lambda$6379.3 is a stellar feature and should be ignored.

We interpret the huge ETR observed for the three DIBs on the left in Fig.~3 as reflecting the population of many high $J$ levels of the large polar carriers of the DIBs, due to radiative pumping by dust emission.  
While this effect is relatively minor for CH$^+$ and CH in which only the lowest excited levels are affected, it can be huge for the DIBs since a great many high $J$ rotational levels are affected because of the low rotational constants of DIB carriers.
For such small rotational constants, the $P$($J$) and $R$($J$) lines are nearly symmetric and linearly spaced by $\sim$~$\pm$2$BJ$ (where $B$ is a rotational constant) at low $J$ and the $Q$($J$) lines pile up at the center of the band.   
As $J$ increases, however, the negative quadratic term ($B'~-~B$)$J(J+1)$ (see next section) takes over and causes an $R$-branch band-head and the ETR.  
The rotational constant in the excited state $B'$ is almost always smaller than that in the ground state $B$ since an electronic excitation tends to weaken chemical bonds, which makes the molecule larger and the moment of inertia higher.

The pronounced ETRs observed for the three DIBs, $\lambda$5780.5, $\lambda$5797.1, and $\lambda$6613.6 indicate that their carriers are polar molecules with dipole moments.  
The absence of ETR for $\lambda$5849.8, $\lambda$6196.0, and $\lambda$6379.3 suggests that either the carriers of those DIBs are non-polar or they are large polar molecules with very small $B'~-~B$.  
Either way, the qualitative difference between the variations of $\lambda$6613.6 and $\lambda$6196.0 eliminates the possibility that they are due to the same molecule, in spite of their near perfect correlation (McCall et al. 2010).

\subsection{Simulation of DIB Profiles with High $T_\mathrm{r}$ and 2.73~K Cosmic Blackbody Radiation}

\subsubsection{Radiative and Kinetic Temperature, $T_\mathrm{r}$ and $T_\mathrm{k}$}

The distribution of molecules in rotational levels is determined as a result of their interaction with the environment through radiation and collision.  
They are influenced by the radiative temperature $T_\mathrm{r}$ of Planck's radiation formula (Planck 1901) and the kinetic temperature $T_\mathrm{k}$ of Maxwell's velocity distribution (Maxwell 1860).  
If $T_\mathrm{r}$~=~$T_\mathrm{k}$, molecules are in equilibrium, following the thermal Boltzmann distribution (Boltzmann 1871) $T_\mathrm{ex}$~=~$T_\mathrm{r}$~=~$T_\mathrm{k}$.  
In interstellar space, the two temperatures are generally different, so a non-thermal molecular distribution is more the rule than the exception.  
Very approximate calculations for simple molecules (e.g. for H$_3^+$, by Oka \& Epp 2004) have been reported, but a calculation for general molecules is beyond the scope of this paper.  
Here we consider the simplest case of linear molecules, based on an approximate treatment of collisions, in order just to see if we can explain the pronounced ETR using realistic molecular parameters. 
Long linear molecules, more specifically acetylenic carbon chain molecules, have been proposed as candidates for carriers of DIBs (Douglas 1977; Maier et al. 2004).  
Many such molecules are known to have large dipole moments, e. g., 3.6~Debye for HC$_3$N (Westenberg \& Wilson), 5.6~Debye for HC$_9$N (Broten et al. 1978) and 11.9~Debye for C$_8$H$^-$ (Br\"{u}nken et al. 2007), and their rotational distributions are sensitive to $T_\mathrm{r}$.   
The result of this treatment is also applicable (approximately) to nearly linear molecules with a dipole moment along the linear axis, such as cumulene carbon chains H$_2$C$_n$ ($n~\geq$~3), which have also been considered as carriers of DIBs (McCarthy et al. 1997).  
Calculations for general molecules are more complicated because of the three (instead of one) rotational constants, quantum numbers, and components of dipole moments, and the lack of spectroscopic and collisional data.  
We will keep that as a future project.
Nevertheless, the results from our simple case give insight for other more complicated molecules as well.

The first excited levels of light molecules like CH$^+$ and CH (40.1~K and 25.6~K above the ground level, respectively) have very short lifetimes for spontaneous emission and their rotational distributions are determined solely by $T_\mathrm{r}$.   
For the large molecules which cause DIBs, the rotational constants are smaller by a few orders of magnitude.  
Therefore, their distributions are governed by collisions (with $T_\mathrm{k}$) for low $J$ rotational levels.  
As $J$ increases, the rate of spontaneous emission increases rapidly ($\propto~J^3$), making radiative effects competitive with collisional effects until the radiative effects (with $T_\mathrm{r}$) take over and dominate at high $J$ levels.  
For example, for HC$_5$N (Avery et al. 1976) with rotational constant 1331.3313~MHz and dipole moment 4.33~Debye, the collisional effects dominate for $J$~=~1, become comparable to the radiative effects at $J~\sim~4$, and the radiative effects dominate for high $J$ levels.  
In the following, this subtle balance between radiative and collisional effects is considered.  
The radiative effects are accurately calculable while the collisional effects are treated approximately.  
In treating the latter, we use the principle of detailed balancing (Boltzmann 1872) as the guiding principle, as in Oka \& Epp (2004).

\subsubsection{Rotational Distribution by Collisions Alone}

In the laboratory where the number density is high, spontaneous emission between rotational levels, $J~\rightarrow~J-1$ is many orders of magnitude slower than collisional processes and is negligible.  
Then the detailed balancing between $J$ and $J$~-~1 levels are expressed as,

\begin{equation}
n(J)C_{J-1}^\downarrow~=~n(J-1)C_{J-1}^\uparrow,
\end{equation}                  ,

\noindent 
where $n(J)$ is the number density of molecules with rotational quantum number $J$, and $C_{J-1}^\downarrow$ and $C_{J-1}^\uparrow$ are rates for collision-induced rotational transitions (Oka 1973) $J~\rightarrow~~J~-~1$ and $J~\leftarrow~J~-~1$, respectively. 
Throughout this paper we follow the spectroscopists' convention, in which the quantum states of higher energy are on the left and exothermic and endothermic processes are expressed by arrows pointing to the right and left, respectively.  
The rates satisfy the principle of detailed balancing, that is,

\begin{equation}
\frac{C_{J-1}^\uparrow}{C_{J-1}^\downarrow}\ = \frac{n(J)_e}{n(J-1)_e}\ = \frac{g_J}{g_{J-1}}\exp(E_{J-1}-E_J)/kT_k = \frac{2J+1}{2J-1}\exp(-2hBJ/kT_k),
\end{equation}

\noindent 
where $n(J)_e$ is the number density in thermal equilibrium, $g_J~=~(2J + 1)$ is the degeneracy of the level, and $E_J~=~hBJ(J + 1)$ is the rotational energy of the molecule in the $J$ level (Townes \& Schawlow 1955).   
From Eqs.~(1) and (2), we can derive the thermal Boltzmann number density with collisions alone as,

\begin{equation}
n(J)~=~n(0)\prod_{m=1}^{J}\frac{C_{m-1}^\uparrow}{C_{m-1}^\downarrow} = \frac{NhB}{kT_k}\ (2J+1)\exp\left[-\frac{hBJ(J+1)}{kT_k}\right],
\end{equation}

\noindent 
after the normalization $n(0)~=~NhB/kT_k$ , where $N$ is the total number of molecules.

\subsubsection{Rotational distribution by collisions and radiation}

When radiation is included along with collisions, the detailed balancing takes the form,

\begin{equation}
n(J)(A^J+B_{J-1}^\downarrow\rho+C_{J-1}^\downarrow) = n(J-1)(B_{J-1}^\uparrow\rho+C_{J-1}^\uparrow),
\end{equation}

\noindent 
where the Einstein $A$ and $B$ coefficients express spontaneous emission and induced radiative effect, respectively (Einstein 1916).  
Using Einstein's relations $A^J~=~(8\pi h\nu^3/c^3)B_{J-1}^\downarrow$, $B_{J-1}^\uparrow/B_{J-1}^\downarrow = (2J+1)/(2J-1)$, and Planck's formula (1901), $\rho~=~8\pi h\nu^3/(e^{h\nu/kT_r}-1)$, we obtain,

\begin{equation}
n(J)\left[A^J\left(1+\frac{1}{e^{h\nu/kT_r}-1}\right) + C_{J-1}^\downarrow \right] = n(J-1)\left[A^J\frac{2J+1}{2J-1}\frac{1}{e^{h\nu/kT_r}-1}+C_{J-1}^\uparrow \right],
\end{equation}

\noindent 
instead of Eq.~1.  
A similar equation was used by Goldreich \& Kwan (1974) for analyzing the radiation trapping by CO.
In the parentheses after $A^J$ on the left hand side of the equation, the first term represents the number of zero point photons per phase space, which causes spontaneous emission (Dirac 1927), and the second term the number of photons (bosons), which causes induced emission and absorption (Landau \& Lifshitz 1969).  
For the $J~\rightarrow~J - 1$ transition, $\nu~=~2BJ$ and,

\begin{equation}
A^J~=~\frac{2^9\pi^4 B^3\mu^2}{3hc^3}\frac{J^4}{2J+1},
\end{equation}

\noindent 
where $\mu$ is the permanent dipole moment.

For collision rates we use approximate formulae (Oka \& Epp 2004),

\begin{equation}
C_{J-1}^\downarrow = C\sqrt{\frac{2J-1}{2J+1}}\exp(hBJ/kT_k), ~~C_{J-1}^\uparrow = C\sqrt{\frac{2J+1}{2J-1}}\exp(-hBJ/kT_k),
\end{equation}

\noindent 
which satisfy the detailed balancing of Eq.~2, where $C$ is a collision rate (independent of $J$) and is proportional to the number density of the environment $n$.  
Using Eq.~5, together with Eqs.~6 and 7, we have,

\begin{equation}
n(J) = n(0)\prod_{m=1}^{J}\left[ \frac{\alpha B^3\mu^2\frac{m^4}{2m-1}\frac{1}{e^{2hBm/kT_r} - 1} + C\sqrt{\frac{2m+1}{2m-1}}e^{-hBm/kT_k}}{\alpha B^3\mu^2\frac{m^4}{2m+1}(1+\frac{1}{e^{2hBm/kT_r} - 1}) + C\sqrt{\frac{2m-1}{2m+1}}e^{hBm/kT_k}}\right],
\end{equation}

\noindent 
where $\alpha~\equiv~2^9\pi^4/3hc^3$.  
Unlike Eq.~3, $n(J)$ depends on both $T_\mathrm{r}$ and $T_\mathrm{k}$, as expected. 
One obvious shortcoming of this approximate treatment of the collisions is that the effect of collision-induced transitions with $\mid\Delta J\mid > 1$ is not included.  
Actually this formalism is a way around dealing with the $\mid\Delta J\mid > 1$ collision induced transitions for which the rates are not known.
We believe that this omission is somewhat justified from the treatment in the previous Section 4.2.2., where such transitions are also not included.  
Note Eqs.~1, 2, and 3 are parallel to Eqs.~4, 5, and 8.

The number densities $n(J)$ in Eq.~8 cannot be expressed analytically as in Eq.~3, so we calculate them numerically.  
They are calculated as a function of the radiative temperatures $T_\mathrm{r}$ due to the dust emission and 2.73~K due to the cosmic blackbody radiation, the kinetic temperature $T_\mathrm{k}$, the rotational constant $B$, the dipole moment $\mu$, and the collision rate $C$.  
As examples, calculated population fractions for $B$~=~1000~MHz,  $\mu$~=~4~Debye, $T_\mathrm{k}$~=~100~K, and $C~=~10^{-7}$~s$^{-1}$ are given in Fig.~4 for $T_\mathrm{r}$~=~2.73~K, 14.6~K, and 80~K.  
As $T_\mathrm{r}$ increases, the molecular populations in higher $J$ levels increase rapidly.

\subsubsection{Simulated Spectra}

The rotational energy of a linear molecule is given by $E_{rot}/h~=~BJ(J + 1)$.  
We ignore small higher order effects such as centrifugal distortion and $\lambda$~~-~doubling.  
The line positions $\nu$ and the H\"{o}nl-London factors $S$ are given as,

\begin{mathletters}
\begin{eqnarray}
R(J):\hspace{0.1in}   \nu = \nu_0 + 2B'(J + 1) + (B' - B)J(J + 1) \hspace{0.3in}        S_\parallel =\frac{J+1}{2J+1} \hspace{0.1in} S_\perp=\frac{J+2}{2(2J+1)} \hspace{0.05in}\\
Q(J): \hspace{0.1in}  \nu = \nu_0 + (B' - B)J(J + 1)     \hspace{1.25in}                 S_\parallel =0   \hspace{0.53in}          S_\perp=1/2   \hspace{0.5in}        \\
P(J): \hspace{0.1in}  \nu = \nu_0 - 2B'J  + (B' - B)J(J + 1)     \hspace{0.75in}         S_\parallel =\frac{J}{2J+1}   \hspace{0.1in}   S_\perp=\frac{J-1}{2(2J+1)},
\end{eqnarray}
\end{mathletters}

\noindent 
where $B'$ is the rotational constant in the upper state of the transition and $S_\parallel$  and $S_\perp$  are the H\"{o}nl-London factors for parallel ($\Delta\lambda=0$) and perpendicular ($\Delta\lambda=\pm1$) transitions, respectively (Herzberg 1989a).
Those factors are for transitions with a singlet state $\lambda=0$ in the lower level, but since $J$ is high for the carriers of DIBs, they give good approximations also for levels with small $\lambda$.
Spectra are calculated by using Eqs.~9 and the populations $n(J)$ calculated in the previous section.   
In order to take into account uncertainty broadening due to a short emission life time of the upper state or internal conversion by the Douglas effect (Douglas 1966), spectral lines are broadened by $\mathit\Gamma~=~1/2\pi\Delta t$.
The simulated profiles are then convolved with the FEROS instrumental function (assumed to be a Gaussian with FWHM~=~6.25~km~s$^{-1}$).

The most crucial molecular parameter for producing the observed ETR is the difference between rotational constants $B' - B$.  
In order to explain the pronounced ETR, we need molecules with large values of $\beta~=~(B' - B)/B$.  
The experimental values of this constant for carbon molecules (mostly from Maier's group) are listed in Table~2.

We note that for all molecules except one (C$_5$) the rotational constant decreases upon electronic excitation, as expected.  
The reported positive change for C$_5$ (Motylewski et al. 1999) is probably due to a mistake in the analysis (J. P. Maier 2013, private communication). 
The large $\beta$ values of 4.3\%, 3.4\%, 5.7\%, 4.9\% etc. are the essence of the observed pronounced ETR.  
A small increase on the order of 1\% cannot reproduce the pronounced ETR unless $T_\mathrm{r}$ is very high.  
This indicates that the carriers of DIBs with pronounced ETR are perhaps relatively small molecules.

The calculated velocity profiles for $\lambda$5780.5, $\lambda$5797.1, and $\lambda$6613.6 are shown in Fig.~5, together with parameters which best mimic the observed ETRs.  
It has been noted in our simulations that, of the seven parameters used in the analysis ($T_\mathrm{r}$, $T_\mathrm{k}$, $B$, $\mu$, $C$, $\beta$, and $\mathit\Gamma$), those related to collisions ($T_\mathrm{k}$ and $C$) are not very critical.  
Even the values of $B$ and $\mu$ are not that critical. 
The most crucial parameters are $T_\mathrm{r}$, $\beta$, and $\mathit\Gamma$ ($\mathit\Delta t$).

\section{DISCUSSIONS}

The essence of the observations and the analysis in this paper is the prominent ETR for the three DIBs shown in the left column of Fig. 3, in sharp contrast with the three DIBs on the right. 
The first author has developed the above formalism based on the premise that the ETRs result from a single ``local'' cloud near Her 36 SE.  
It is possible instead that the DIBs toward Her 36 are a composite of those local DIBs and (as seen toward 9 Sgr) foreground DIBs, and that the latter need to be subtracted.  
In that view, the observed profiles reflect an accidental blend of DIBs from two clouds:  one in the foreground (without ETR) and the other local to Herschel 36.
The foreground-subtracted profiles shown (in black) in the left-hand panel of Fig. 8 of Paper I are difficult to understand, spectroscopically, however, as the usually observed DIB cores are nearly absent.
On the other hand, if the foreground absorption were significantly weaker toward Herschel 36 than toward 9 Sgr (as shown in the right-hand panel of Fig. 8 of paper I), then the subtracted profiles would appear to be more consistent with the analysis given above. 
That analysis and the discussion and conclusions given below are based on the premise that the beautiful and systematic ETRs, which make sense spectroscopically, must be due to a single, local cloud. 

\subsection{$T_\mathrm{r}$ and $\beta$}

The above analysis has indicated that a large fractional decrease in the rotational constant $\beta$ ($\sim$~5\%) and a high radiative temperature ($T_\mathrm{r}$ $>$~50~K) in the environment of the DIB carriers are needed to produce the observed pronounced ETRs.
$T_\mathrm{r}$~=14.6~K is not high enough to explain the ETR.  
The width of an ETR is simply proportional to $\beta$, and high $T_\mathrm{r}$ extends the rotational quantum number $J$ of the populated levels (as seen in Fig.~4) and thus greatly extends the ETR. 
The requirement of such a high $T_\mathrm{r}$ is surprising, but there seems to be no alternative, since the value of $\beta$ is limited (Table 2).  
Our calculations are only for linear molecules, but it seems that this requirement holds for other types of molecules with three rotational constants, $A$, $B$, and $C$, as long as the ETRs are ascribed to the differences between the rotational constants of the excited and ground states.  
Other possible mechanisms, albeit not very realistic, are discussed below in Sec. 5.3.

\subsection{$\lambda$6613.6}

Although our model calculations for polar linear molecules with high $\beta$ and $T_\mathrm{r}$ give the observed pronounced ETR for the three DIBs as shown in Fig.~5, we cannot claim that they are the carriers of the DIBs.  
In particular, a linear molecule cannot explain the triplet structure of DIB $\lambda$6613.6 discovered by Sarre et al. (1995).  
A perpendicular transition of a linear molecule gives a triplet structure but the central pileup of the $Q$ branch lines should be stronger than those of the $P$ and $R$ branches by a factor of approximately 2 due to the H\"{o}nl-London factors -- which contradicts the observation.  
The claim by Schultz et al. (2000) that $\lambda$6613.6 is due to a parallel band of a long chain cumulane molecule (H$_2$C$_n$) does not work either, since the $Q$ branch pileup is weak for such molecule.

Kerr et al. (1996) proposed a large planar symmetric top as the carrier of this band.  
Such a molecule does not have a dipole moment and thus cannot be the carrier of $\lambda$6613.6 (which is sensitive to $T_\mathrm{r}$).  
In general, non-polar molecules often considered for DIBs (such as C$_n$, HC$_n$H, H$_2$C$_n$H$_2$, NC$_n$N, etc.), symmetric PAHs (like pyrene, coronene, ovalene etc.), or C$_{60}$ and their cations and anions cannot be the carriers of  DIBs exhibiting the ETR.  
Derivatives of PAHs, polar PAHs such as phenanthrene, tetraphen, pentaphene or recently synthesized olympicene could be the carriers, but most of them probably do not have sufficiently large $\beta$ to produce the ETR.
There are no experimental data for $\beta$ for coronene, but a theoretical value $\beta$~=~0.3\% (Cossart-Magos \& Leach 1990) is an order of magnitude too small to produce the pronounced ETR.

Cami et al. (2004) discovered small variations in the structure of $\lambda$6613.6 for different sight lines and interpreted them as due to variations in rotational temperature from 21~K to 25~K.  
We believe that the rotational temperature is much lower (because of spontaneous emission), and a minute increase of $T_\mathrm{r}$ over the 2.73 K blackbody radiation is sufficient to explain the shifts (due to excitation of $P$ and $R$ lines of low $J$).

A similar difficulty is encountered also for $\lambda$5797.1.  
The high resolution spectrum of this DIB shows a sharp $Q$ branch line.
Our model calculation requires a large line broadening which is obviously contradictory.  
Here we are just content to be able to produce observed huge ETRs using realistic $\beta$ and $T_\mathrm{r}$ and leave those thorny problems for future studies.

\subsection{Possible Other Interpretations}

We ascribe the pronounced ETR to small ($\sim$~5\%) variations in the rotational constants, but there may be two other possible interpretations.  
It is known that some molecules change their shape from non-linear in the ground electronic state to linear in the excited state.  
Some examples are CH$_2$ (\emph{\~{B}}$^3\Sigma_u^-$~$\leftarrow$~\emph{\~{X}}$^3B_1$), HCO (\emph{\~{A}}$^2\Pi$~$\leftarrow$~\emph{\~{X}}$^1A'$) and NO$_2$ (\emph{\~{E}}$^2\Sigma_u^+$~$\leftarrow$~\emph{\~{X}}$^2A_1$) (Herzberg 1989b).  
These molecules are too small to cause DIBs, but if there are larger molecules with the same property, the change in the rotational constant $A$ by the electronic excitation is 100\% and radiative pumping may cause pronounced ETR without having to assume a high $T_\mathrm{r}$. 
While we have not been able to find relevant experimental data for large molecules of this nature, it is a possibility. 
Such electronic transitions, however, tend to have many vibronic satellites with high Franck-Condon factors which are not seen for $\lambda$5780.5, $\lambda$5797.1, and $\lambda$6613.6.

Alternatively, we may interpret the ETR as a result of the excitation of low frequency vibrational modes (rather than rotational levels in the ground state).  
The carriers then would not need to be polar molecules.  
The radiative effect for a vibrational excitation is typically 100 times less than that for rotational excitation, but the collisional excitations are also small and they may compete. 
The bending frequencies of C$_3$, 63.4~cm$^{-1}$ (Schmuttenmaer et al. 1990), and of C$_3$O$_2$, 18.3~cm$^{-1}$ (Fusina et al. 1980), are too high, but there may be larger molecules with lower vibrational frequencies.  
Such spectra, however, are likely to exhibit blue as well as red wings. 
Also its effect will look more like the three examples on the right of Fig.~3, $\lambda$5849.8, $\lambda$6196.0, and $\lambda$6379.3.

\section{CONCLUSIONS}

We have interpreted the pronounced ETRs observed for $\lambda$5780.5, $\lambda$5797.1, and $\lambda$6613.6 toward Herschel 36 as the result of the variation of the rotational constant upon electronic excitation.
Our results do not lead to identification of the specific carriers of DIBs, but they exclude some possibilities with various degrees of certainty.

\subsection{Firm Conclusions}

The presence or absence of the ETR allows discrimination between polar and non-polar carriers. 
The carriers of $\lambda$5780.5, $\lambda$5797.1, and $\lambda$6613.6, which show pronounced ETR, cannot be non-polar molecules.  
Molecules like pure carbon chains (C$_n$) or symmetric hydrocarbon chains (HC$_n$H, H$_2$C$_n$H$_2$, NC$_n$N, etc.) and their cations and anions cannot be the carriers of those DIBs.  
Likewise, symmetric PAHs benzene, pyrene, coronene, ovalene etc. and their cations and anions (Cossart-Magos \& Leach 1990; see also the many PAHs discussed in Sharp et al. 2006) or C$_{60}$ and their cations and anions cannot be the carriers of DIBs exhibiting strong ETR.

The carriers of DIBs which show the ETR (on the left of Fig.~3) cannot be the carriers of the DIBs without ETR (on the right).  
In particular, the carriers of $\lambda$6196.0 and $\lambda$6613.6 cannot be the same molecule, despite the near-perfect correlation of those two DIBs (McCall et al. 2010).

\subsection{Likely Conclusions}

If linear molecules are assumed (as in Section 4), the number of heavy atoms is likely 3--6 but not much higher, since the values of $\beta$~=~$(B'-B)/B$ are not high enough for heavier molecules to produce the pronounced ETR (unless the radiative temperature is extremely high).
Many linear molecules of this size have already been studied spectroscopically by Maier's group, but cumulene carbene molecules $l$-H$_2$C$_{\rm n}$ (n $>$ 3) may be candidates.
Non-linear molecules are outside the scope of this paper.
The high column densities required to produce the strong DIBs are problems (Oka \& McCall 2011; Liszt et al. 2012), but this will be a problem for any candidate molecules of this size.

The carriers of $\lambda$5849.8, $\lambda$6196.0, and $\lambda$6379.3, which do not show the ETR as observed in $\lambda$5780.5, $\lambda$5797.1, and $\lambda$6613.6, are most likely non-polar molecules.  
The possibility that the carriers of those DIBs exist in different location from other DIBs and are not exposed to the dust emission is perhaps safely neglected.  
However, our simulations have not excluded the possibility that they may be due to large polar molecules with very small $\beta$.

In this paper we have considered only linear molecules where the ETR is created by large $\beta$~=~$(B' - B)/B$ ($\sim$~5\%).  
For non-linear molecules, there are other possibilities, such as large decreases in the differences of rotational constants $A - B$ or $C - B$ for prolate or oblate tops, respectively.  
Either way, the relative variations of these constants are not very likely to be much higher than 5~\%, except for the case mentioned in Section 5.3.
In most cases, they are 10 times smaller than needed to produce the pronounced ETR.

\subsection{Speculation}

Since carriers of DIBs with pronounced ETRs are polar molecules with modest size, they should be observable by radio astronomy.  
Because of the low density and subthermal rotational distribution in diffuse clouds, the observation need to be by absorption (Liszt et al. 2012), and in relatively long wavelength region for low $J$ transitions.

\begin{center}
\textbf{Acknowledgments}
\end{center}

We are indebted to S. Federman and A. Witt for criticisms of this paper.
T. O. acknowledges helpful discussions with H. Liszt, J. P. Maier, and P. Sarre. 
He also thanks J. Luque and D. R. Crosley for providing theoretical calculations on CH.  
We acknowledge support by NSF grants AST 1109014 (TO), AST 1008424 (JD), AST 1009639 (DGY), and AST 1238926 (DEW).

\clearpage

\begin{figure}
\plotone{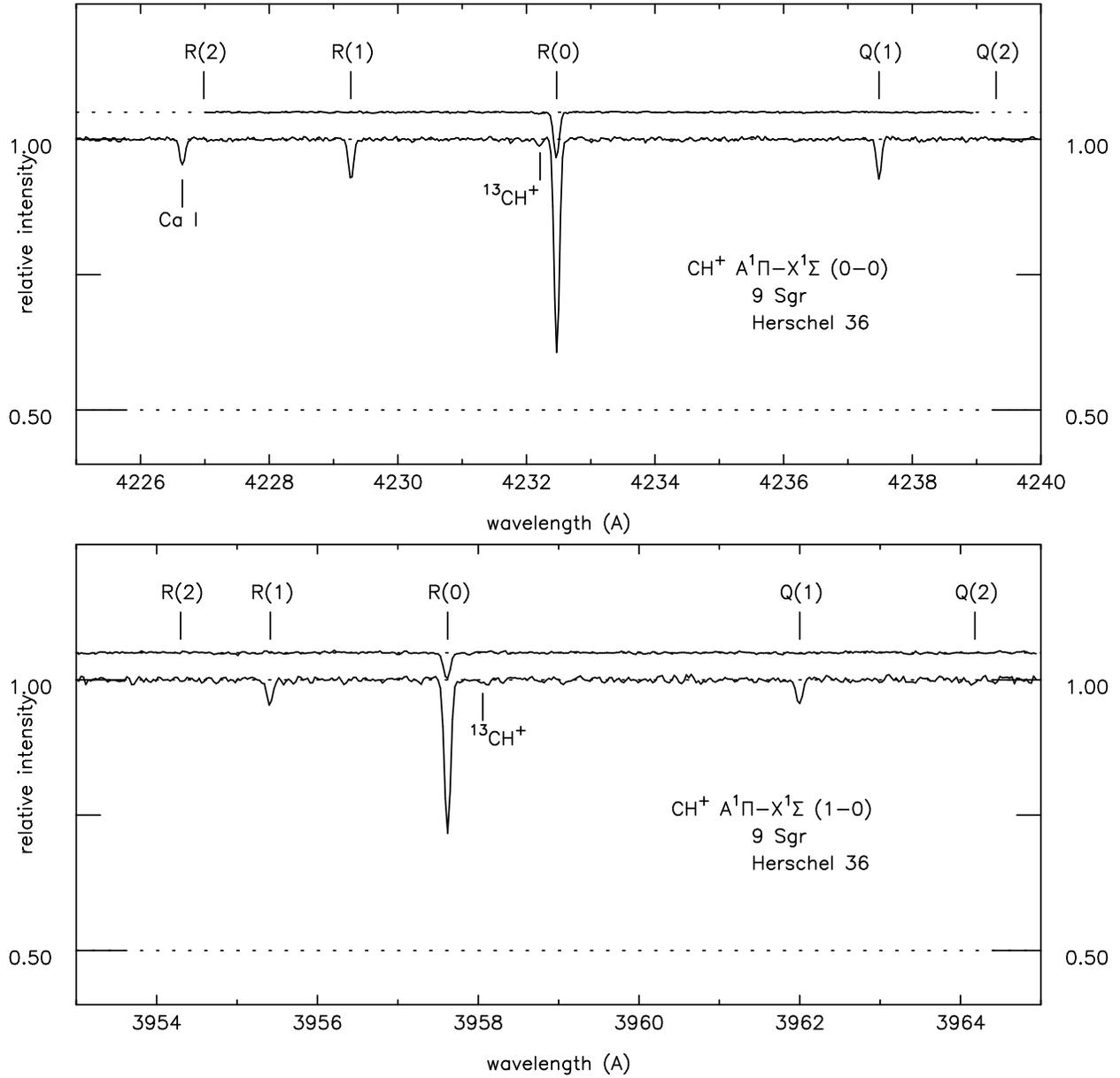}
\caption{The (0,0) band (above) and (1,0) band (below) of the \emph{\~{A}}$^1\Pi$~$\leftarrow$~\emph{\~{X}}$^1\Sigma$ transitions of
CH$^+$ observed with FEROS toward 9 Sgr (upper trace) and Herschel 36 (lower trace).  
The spectral resolution is $R$~=~48,000; wavelengths are heliocentric and in air.}
\label{fig2}
\end{figure}
\clearpage

\begin{figure}
\plotone{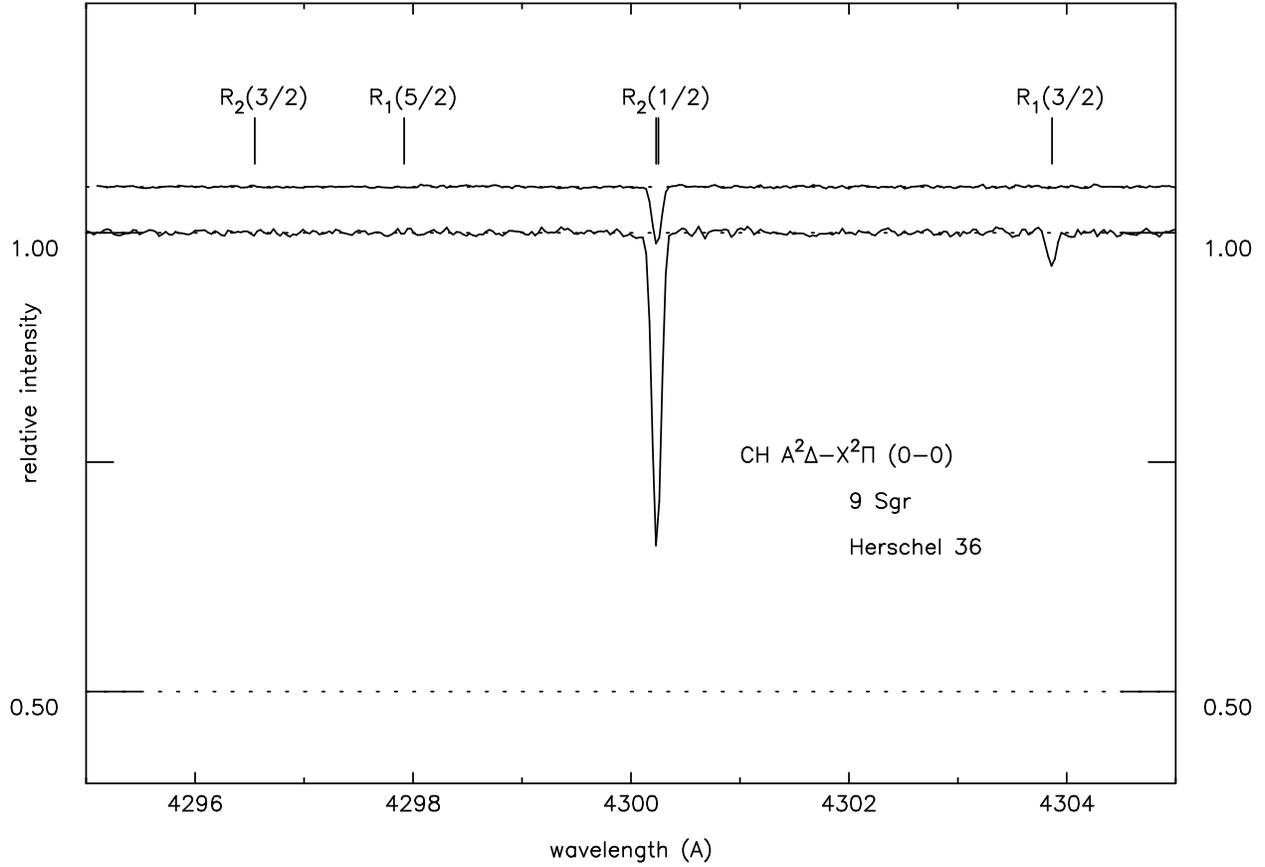}
\caption{The R$_2$(1/2) and R$_1$(3/2) lines of the \emph{\~{A}}$^2\Delta$~$\leftarrow$~\emph{\~{X}}$^2\Pi$ (0,0) transition of CH observed with FEROS toward 9 Sgr (upper trace) and Herschel 36 (lower trace).
Wavelengths are heliocentric and in air.}
\label{fig3}
\end{figure}
\clearpage

\begin{figure}
\plotone{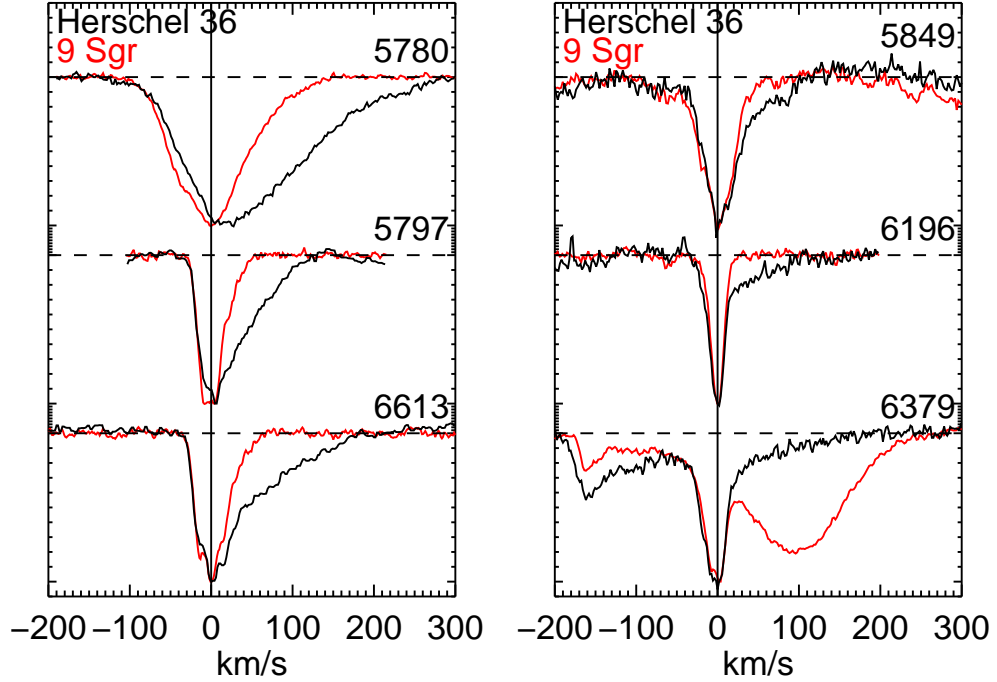}
\caption{A comparison of DIBs $\lambda$5780.5, $\lambda$5797.1, and $\lambda$6613.6 (left) and $\lambda$5849.8, $\lambda$6196.0, and $\lambda$6379.3 (right) toward Her 36 (black) and 9 Sgr (red) observed with FEROS.  
The 9 Sgr profiles have been scaled to the central depth seen toward Her 36; the velocity zero points have been set by the interstellar \ion{K}{1} $\lambda$7698.965 line.
The pronounced extension of tail toward red (ETR) observed toward Her 36 for the former three is due to radiative pumping by dust emission from the adjacent IR source Her 36 SE, indicating that the carriers of those three DIBs are polar molecules.  
The carriers of the latter three DIBs which do not show the strong ETR are likely non-polar molecules.  
The deep broad absorption near $\lambda$6379.3 for 9 Sgr is a stellar absorption line.}
\label{fig4}
\end{figure}
\clearpage

\begin{figure}
\plotone{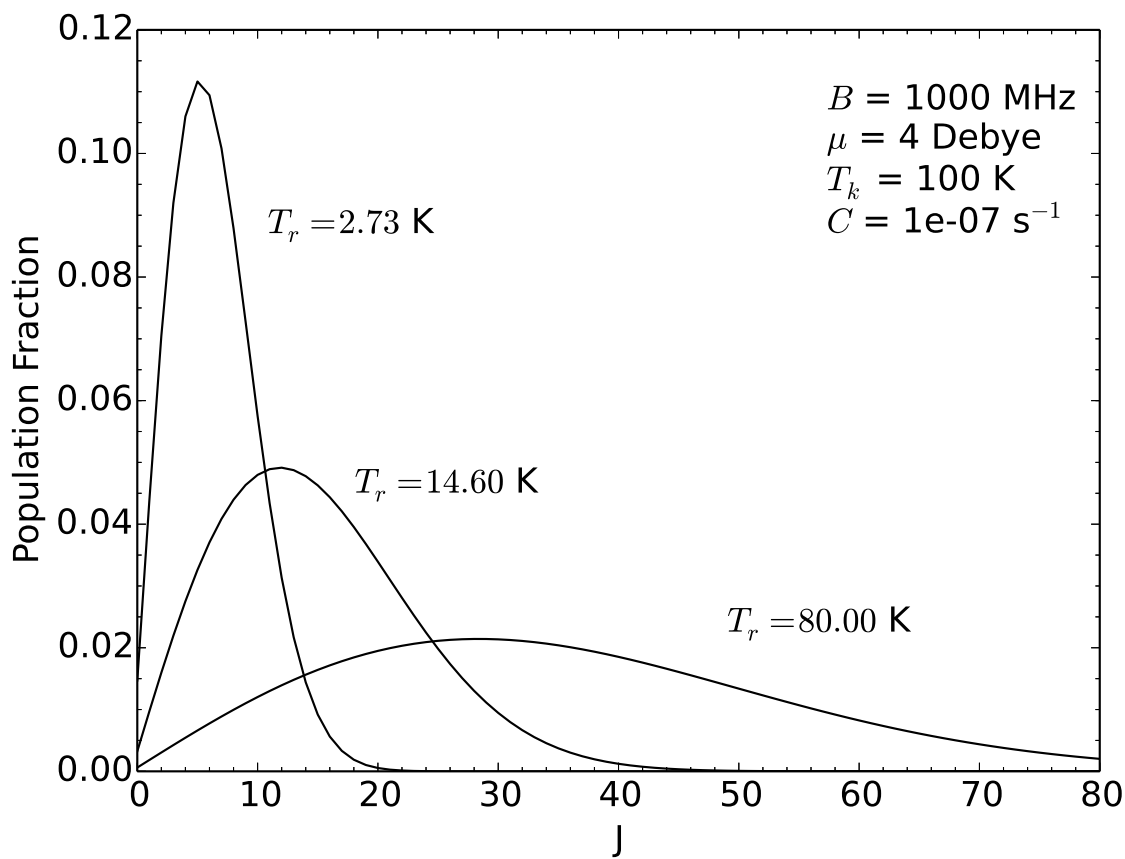}
\caption{Population fractions of polar molecules calculated by using $n(J)$ of Eq.~8 for $T_\mathrm{r}$ = 2.73 K, 14.6 K, and 80 K.  
Spontaneous emission reduces the $n(J)$ in high $J$ levels for low $T_\mathrm{r}$, while the dust emission pumps them back for high $T_\mathrm{r}$.}
\label{fig5}
\end{figure}
\clearpage

\begin{figure}
\epsscale{0.4} 
\plotone{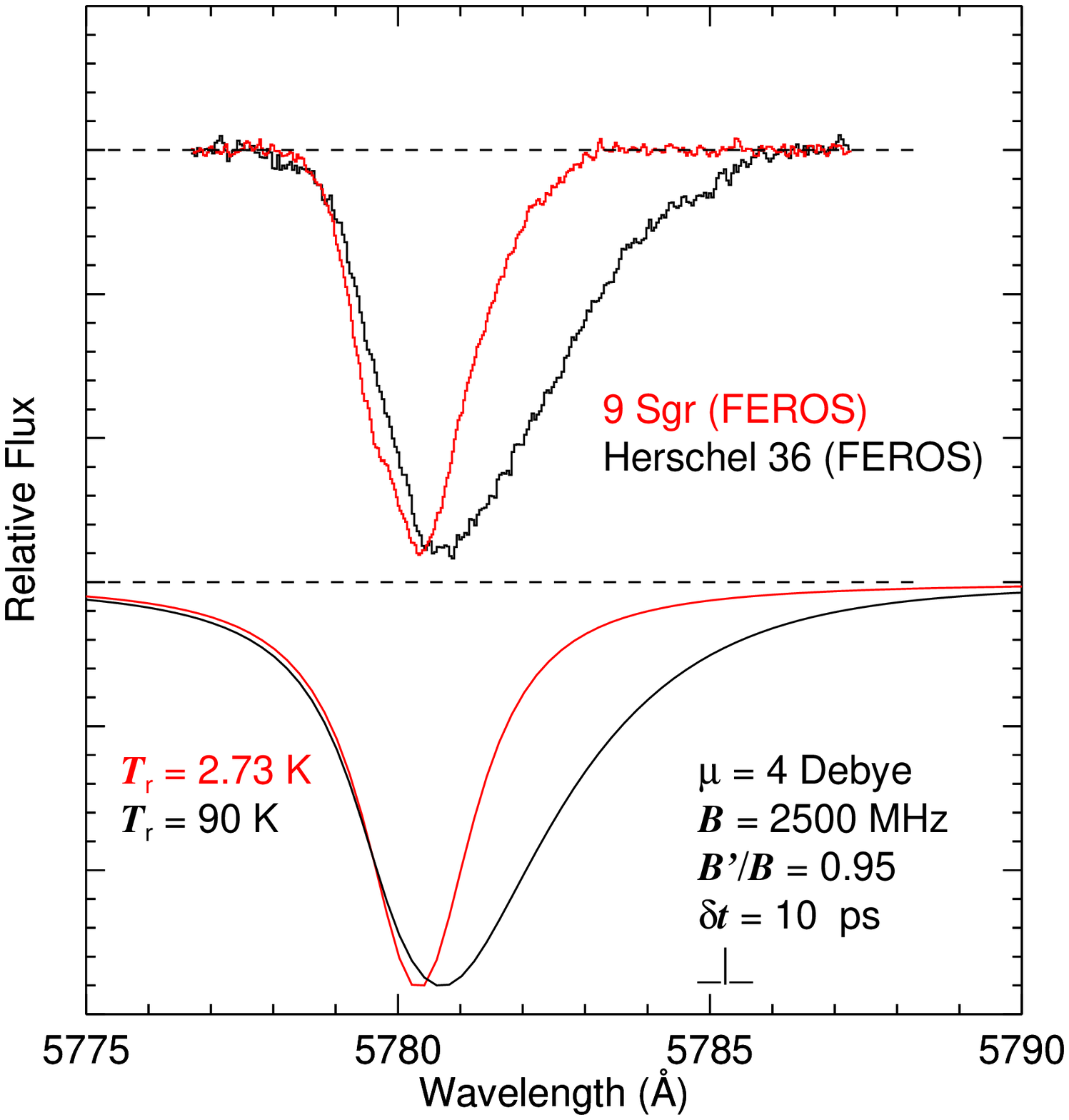}
\plotone{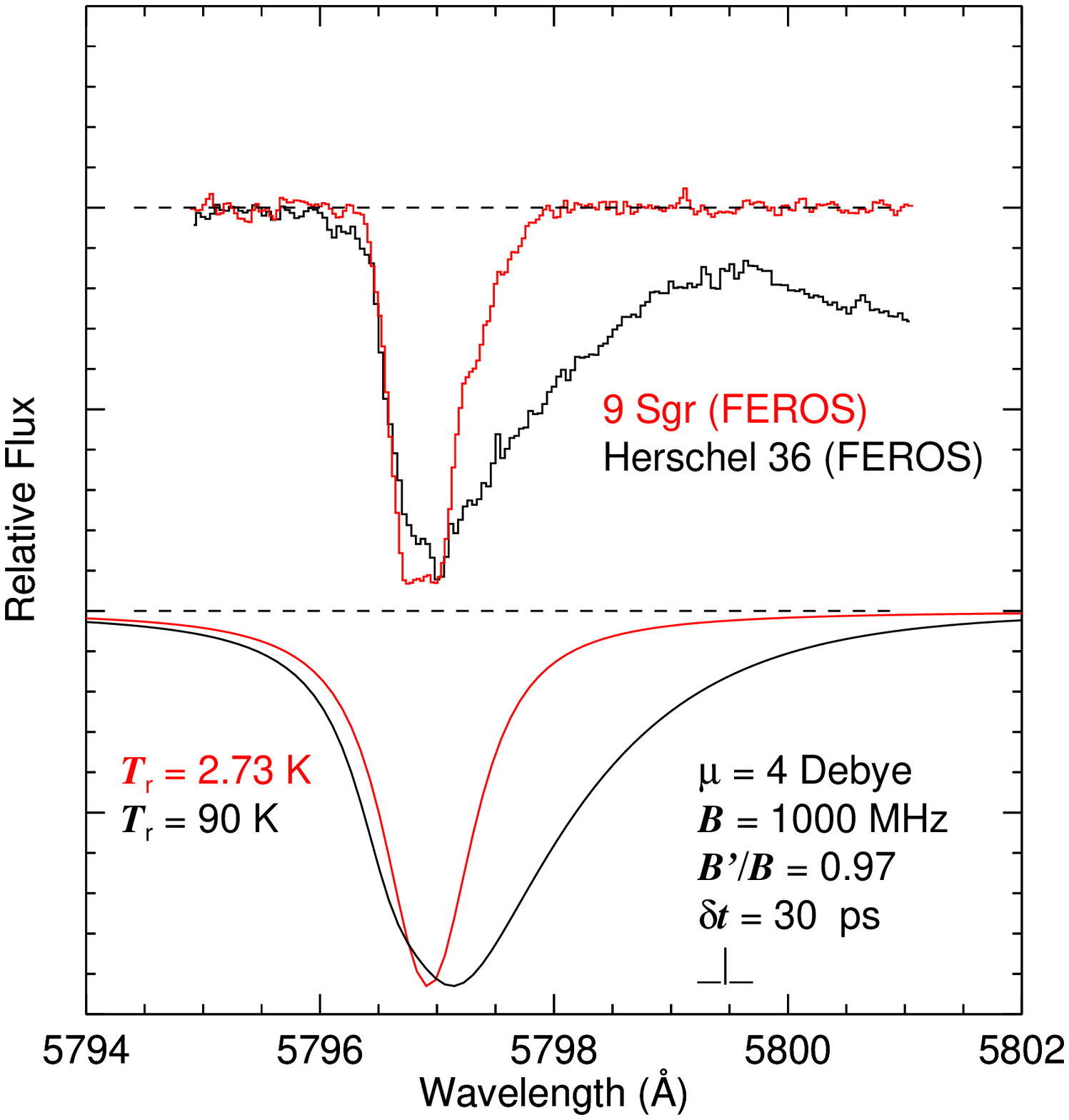}
\plotone{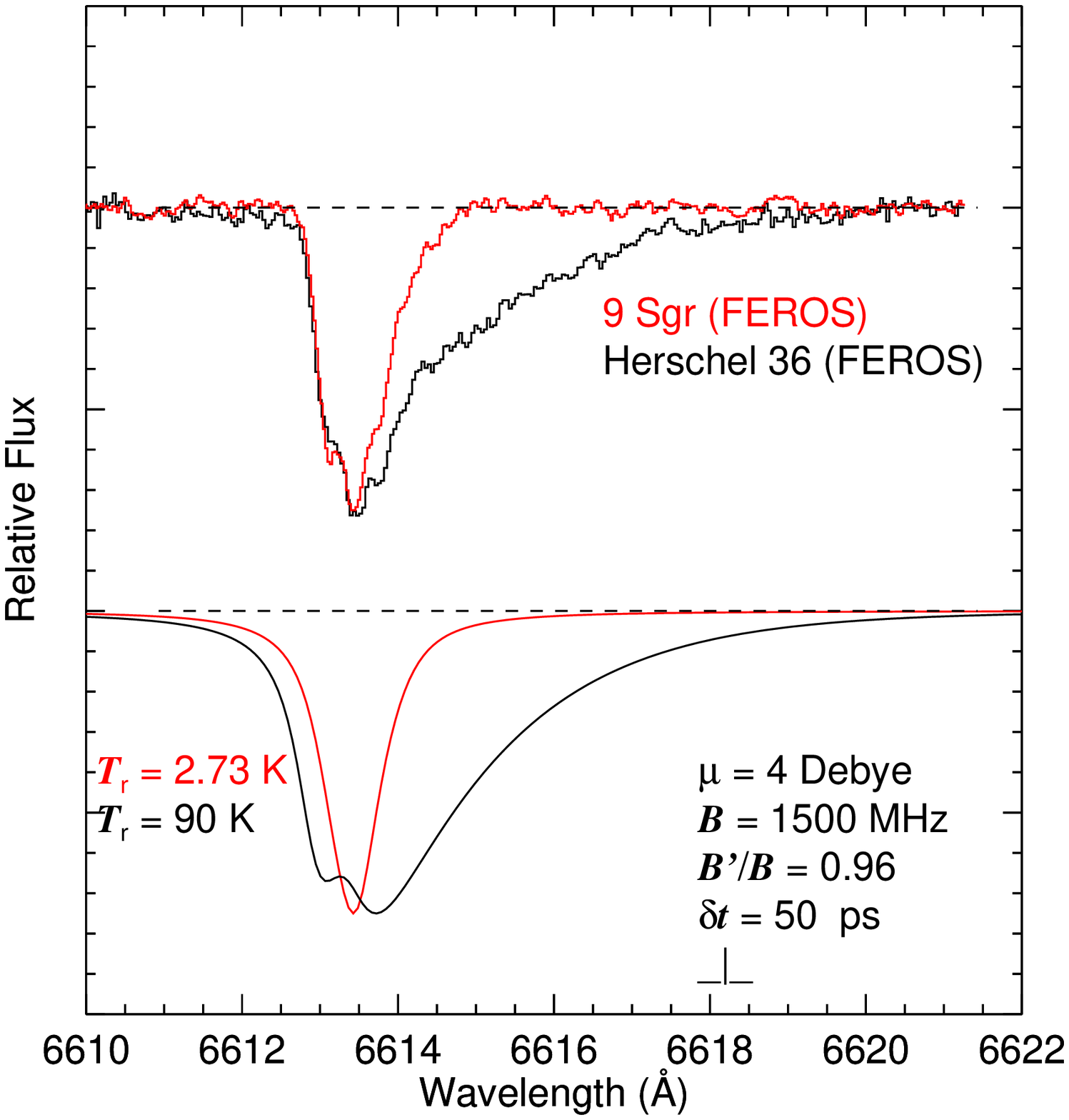}
\caption{A comparison of observed DIBs $\lambda$5780.5, $\lambda$5797.1, and $\lambda$6613.6 toward Herschel 36 (black) and 9 Sgr (upper traces) and the results of model calculations (lower traces).  
The 9 Sgr and model profiles have been scaled to the central depth seen toward Her 36.
NOTE: the values of $\delta t$ should be divided by 2$\pi$.}
\label{fig6}
\end{figure}

\clearpage

\begin{deluxetable}{cccccccc}
\tabletypesize{\scriptsize}
\tablecaption{Observed transitions of CH$^+$ and CH toward Herschel 36, with oscillator strengths $f$, observed wavelengths (in air) $\lambda_{air}$, equivalent widths $W_\lambda$, H\"{o}nl-London factors $S$, derived column densities $N(J)$, and excitation temperatures $T_\mathrm{ex}$. \label{tbl-1}}
\tablewidth{0pt}
\tablehead{
\colhead{Molecule} &
\colhead{Transition} &
\colhead{Line} &
\colhead{$\lambda_\mathrm{air}$} &
\colhead{$W_\lambda$} &
\colhead{S} &
\colhead{$N(J)$} &
\colhead{$T_\mathrm{ex}$}\\
\colhead{ } &
\colhead{ } &
\colhead{ } &
\colhead{(\AA)} &
\colhead{(m\AA)} &
\colhead{ } &
\colhead{(10$^{13}$cm$^{-2}$)} &
\colhead{(K)}}

\startdata
CH$^+$ & \emph{\~{A}}$^2\Delta$~$\leftarrow$~\emph{\~{X}}$^2\Pi$(0,0)$^a$        & $R(2)$     & 4226.982 & $<$~0.9      & 0.4 & $<$~0.26    & $<$~22.9    \\
       & $f$~=~0.00545$^{b}$                                                     & $R(1)$     & 4229.273 &  7.3$\pm$0.3 & 0.5 & 1.8$\pm$0.1 & 14.3$\pm$0.5\\
       &                                                                         & $R(0)$     & 4232.472 & 40.2$\pm$0.3 & 1   & 9.9$\pm$0.6 &             \\
       &                                                                         & $Q(1)$     & 4237.481 &  6.6$\pm$0.3 & 0.5 & 1.8$\pm$0.1 & 14.3$\pm$0.5\\
       &                                                                         & $Q(2)$     & 4239.302 & $<$~0.9      & 0.5 & $<$0.26     & $<$~22.9    \\
$^{13}$CH$^+$ & \emph{\~{A}}$^2\Delta$~$\leftarrow$~\emph{\~{X}}$^2\Pi$(0,0)$^c$ & $R(0)$     & 4232.200 &  1.1$\pm$0.3 & 1   &0.14$\pm$0.02&             \\
 & \\
CH$^+$ & \emph{\~{A}}$^2\Delta$~$\leftarrow$~\emph{\~{X}}$^2\Pi$(1,0)$^a$        & $R(1)$     & 3955.407 &  4.4$\pm$0.5 & 0.5 & 1.9$\pm$0.1 & 15.4$\pm$0.4\\
       & $f$~=~0.00331$^{b}$                                                     & $R(0)$     & 3957.618 & 27.0$\pm$0.5 & 1   & 8.6$\pm$0.4 &             \\
       &                                                                         & $Q(1)$     & 3961.994 &  4.2$\pm$0.5 & 0.5 & 1.9$\pm$0.1 & 15.4$\pm$0.4\\
$^{13}$CH$^+$ & \emph{\~{A}}$^2\Delta$~$\leftarrow$~\emph{\~{X}}$^2\Pi$(1,0)$^a$ & $R(0)$     & 3958.100 &  1.0$\pm$0.4 & 1   &0.19$\pm$0.04&             \\
 & \\
CH     & \emph{\~{A}}$^2\Delta$~$\leftarrow$~\emph{\~{X}}$^2\Pi$(0,0)$^d$        & $R_2(1/2)$ & 4300.227 & 36.0$\pm$0.4 & 1   & 6.9$\pm$0.2 &             \\
       & $f$~=~0.00505$^{e}$                                                     & $R_1(3/2)$ & 4303.861 &  3.4$\pm$0.3 & 1.5 &0.30$\pm$0.02&  6.7$\pm$0.1\\
CH     & \emph{\~{B}}$^2\Sigma^{-}$~$\leftarrow$~\emph{\~{X}}$^2\Pi$(0,0)$^f$    & $R_2(1/2)$ & 3878.700 &  5.5$\pm$0.5 & 1/3 & 3.8$\pm$0.2 &             \\
       & $f$~=~0.00320$^{g}$                                                     & $Q_2(1/2)$ & 3886.341 & 12.2$\pm$0.6 & 1   & 3.3$\pm$0.1 &             \\
       &                                                                         & $P_2(1/2)$ & 3890.148 &  8.5$\pm$0.8 & 2/3 & 3.2$\pm$0.2 &             \\
\enddata
\tablerefs{$^a$ Carrington \& Ramsay (1982); $^b$Larsson \& Siegbahn (1983a); $^c$Bembenek et al. (1997); $^d$Zachwieja (1995);
$^e$Larsson \& Siegbahn (1983b); $^f$K\c{e}pa et al. (1996); $^g$Lien (1984)}
\end{deluxetable}

\clearpage

\begin{deluxetable}{cccccc}
\tabletypesize{\scriptsize}
\tablecaption{Variation of the rotational constants of carbon molecules upon electronic excitation. \label{tbl-2}}
\tablewidth{0pt}
\tablehead{
\colhead{Molecule} &
\colhead{Transition} &
\colhead{$B$} &
\colhead{$B'-B$} &
\colhead{$\beta=(B'-B)/B$} &
\colhead{Reference}\\
\colhead{ } &
\colhead{ } &
\colhead{(cm$^{-1}$)} &
\colhead{(cm$^{-1}$)} &
\colhead{(\%)} &
\colhead{ }}

\startdata
C$_3$             & $^1\Pi_u$~$\leftarrow$~$^1\Sigma_g^+$      & 0.4305       & -0.0183       & -4.3     & c \\
$\ell$-H$_2$C$_3$ & $^1A_1$~$\leftarrow$~$^1A_2$               & 0.346783$^a$ & -0.01182$^a$  & -3.4$^a$ & d \\
C$_4$             & $^3\Sigma_u^-$~$\leftarrow$~$^3\Sigma_g^-$ & 0.166111     & -0.0091       & -5.7     & e \\
HC$_4$H$^+$       & $^2\Pi_u$~$\leftarrow$~$^2\Pi_g$           & 0.14690      & -0.00681      & -4.9     & f \\
C$_5$             & $^1\Pi_u$~$\leftarrow$~$^1\Sigma_g^+$      & 0.0853133    &  0.0002$^b$   &  0.2$^b$ & g \\
HC$_6$H$^+$       & $^2\Pi_u$~$\leftarrow$~$^2\Pi_g$           & 0.0445943    & -0.0008022    & -1.8     & h \\
HC$_8$H$^+$       & $^2\Pi_u$~$\leftarrow$~$^2\Pi_g$           & 0.0190779    & -0.0002106    & -1.1     & i \\
HC$_7$N$^+$       & $^2\Pi$~$\leftarrow$~$^2\Pi$               & 0.0189665    & -0.001934     & -1.0     & j \\
C$_{24}$H$_{12}$  &                                            & 0.011123$^l$ & -0.000033$^l$ & -0.3$^l$ & k \\
\enddata

\tablenotetext{a}{$(B+C)/2$ is used instead of $B$.}
\tablenotetext{b}{The positive change of the rotational constant upon excitation is probably due to a mistake in the analysis (see text).}
\tablerefs{c Gausset et al. (1965); d Achkasova et al. (2006); e Linnartz et al. (2000); f Kuhn et al. (1986); g Motylewski et al. (1999);
h Sinclair et al. (1999); i Pluger et al. (2000); j Sinclair et al. (2000); k Cossart-Magos \& Leach 1990; l Theoretical values}
\end{deluxetable}

\end{document}